\documentclass[letter,twocolumn]{jpsj3}
\usepackage[dvipdfmx]{graphicx}

\makeatletter % bblファイルのエラー回避(Overleaf)
\newcommand\newblock{\hskip .11em\@plus.33em\@minus.07em}
\makeatother

\usepackage{amsmath, amssymb}
\usepackage{bm}
\usepackage{mathrsfs}
\usepackage{physics}
\usepackage{comment}
\usepackage[normalem]{ulem}

\usepackage{color}

\title{Nonrelativistic Functional Properties in Collinear Antiferromagnets Based on Multipole Representation Theory}
\author{Yuuki Ogawa and Satoru Hayami}
\inst{Graduate School of Science, Hokkaido University, Sapporo 060-0810, Japan}

\abst{
In recent years, the concept of multipoles has been widely used to describe and classify various magnetic and electric responses in solids, providing a systematic way to identify symmetry-allowed or -forbidden physical responses.
Conventionally, multipole classifications rely on the magnetic point group of a system, which inherently incorporates the effects of relativistic spin--orbit coupling because the spin orientation is supposed to follow the point-group transformation of the lattice.
However, this approach becomes insufficient in situations where relativistic spin--orbit coupling is negligibly weak or where the spin and orbital (lattice) degrees of freedom are decoupled, thereby requiring a more comprehensive symmetry description.
In this work, we introduce a multipole description on the basis of the spin-point-group symmetries, enabling a systematic exploration of nonrelativistic phenomena that persist even without spin--orbit coupling in a collinear antiferromagnet. 
As an application, we theoretically demonstrate spin-current generation driven by elastic waves in a specific collinear antiferromagnet, fully independent of spin--orbit coupling.
}

\date{\today}

\begin{document}
\maketitle

There has been increasing interest in magnetic and transport phenomena free from the relativistic spin--orbit coupling (SOC), opening broader avenues for spintronics beyond conventional heavy-metal systems~\cite{Naish1962, Kitz1965, Brinkman1966,Litvin_Opechowski1974,Litvin1977, Liu2022 , Watanabe2024, Xiao2024, Chen2024,Jiang2024, Chen2025, SciPostPhys.18.3.109, Etxebarria2025}. 
Such developments have been particularly highlighted in the context of altermagnets~\cite{Noda2016, Okugawa2018, Ahn2019, Naka2019, Hayami_YK2019, Hayami_PhysRevB.102.144441, Yuan2020, Naka_PhysRevB.103.125114, Yuan2021PRMat, Yuan2021PRB, Gonzalez-Hernandez2021, Smejkal2022PRX1, Smejkal2022PRX3, Smejkal2022PRX4}, where collinear antiferromagnets breaking time-reversal ($\mathcal{T}$) symmetry exhibit momentum-dependent spin-split band structures in the form of $k_\alpha k_\beta \sigma$ in $d$-wave type and $k_\alpha k_\beta k_\gamma k_\delta \sigma$ in $g$-wave type ($k_\alpha$ and $\sigma$ represent $\alpha = x, y, z$ component of the wave vector and the spin, respectively) even without the SOC. 
Consequently, altermagnets realize a variety of SOC-free responses, such as spin current generation~\cite{Ahn2019,Naka2019, Hayami2022spinconductivity, Sourounis_PhysRevB.111.134448, Ezawa_PhysRevB.111.125420, Naka2025altermagnetic}, linear and nonlinear piezomagnetic effect~\cite{Ma2021, Zhu2024, Aoyama2024, McClarty_PhysRevLett.132.176702, Wu2024, Yershov2024, Chen2024, Ogawa2025, Huyen2025, Naka2025}, 
nonlinear magnetoelectric effect~\cite{Oike_PhysRevB.110.184407}, and electric polarization (magnetization) under the magnetic (electric) field gradient~\cite{sato2025quantum, oike2025thermodynamic}.

Following the initial recognition in these $\mathcal{T}$-broken collinear antiferromagnets, recent studies have expanded their scope to noncollinear antiferromagnets breaking spatial inversion ($\mathcal{P}$) symmetry~\cite{Hayami_PhysRevB.101.220403, Hayami_PhysRevB.102.144441, Hayami_PhysRevB.105.024413, Brekke_PhysRevLett.133.236703, Sukhachov_PhysRevB.110.205114, Soori_PhysRevB.111.165413, Nagae_PhysRevB.111.174519, Hodt_PhysRevB.111.205416}, which give rise to the antisymmetric spin splitting in the form of $k_\alpha \sigma$ in $p$-wave type and $k_\alpha k_\beta k_\gamma\sigma$ in $f$-wave type. 
Moreover, the exploration of nonrelativistic properties has broadened to $\mathcal{PT}$-symmetric antiferromagnets with the spin degeneracy, yet still allowing spin-dependent phenomena, such as the nonlinear (spin) Hall effect and photocurrent generation~\cite{Hayami_PhysRevB.106.024405, Kondo_PhysRevResearch.4.013186, Yuan2023, Watanabe_Yanase2024, Matsuda2025} 
and $\mathcal{T}$-broken noncoplanar antiferromagnets exhibiting the anomalous Hall effect~\cite{Ohgushi_PhysRevB.62.R6065, Shindou_PhysRevLett.87.116801, Martin_PhysRevLett.101.156402} and nonreciprocal transport~\cite{Hayami_PhysRevB.106.014420}.
These developments have shown that a variety of nonrelativistic responses emerge in a wide range of magnets.

\begin{figure}[t]
    \centering
    \includegraphics[width=1.0\linewidth]{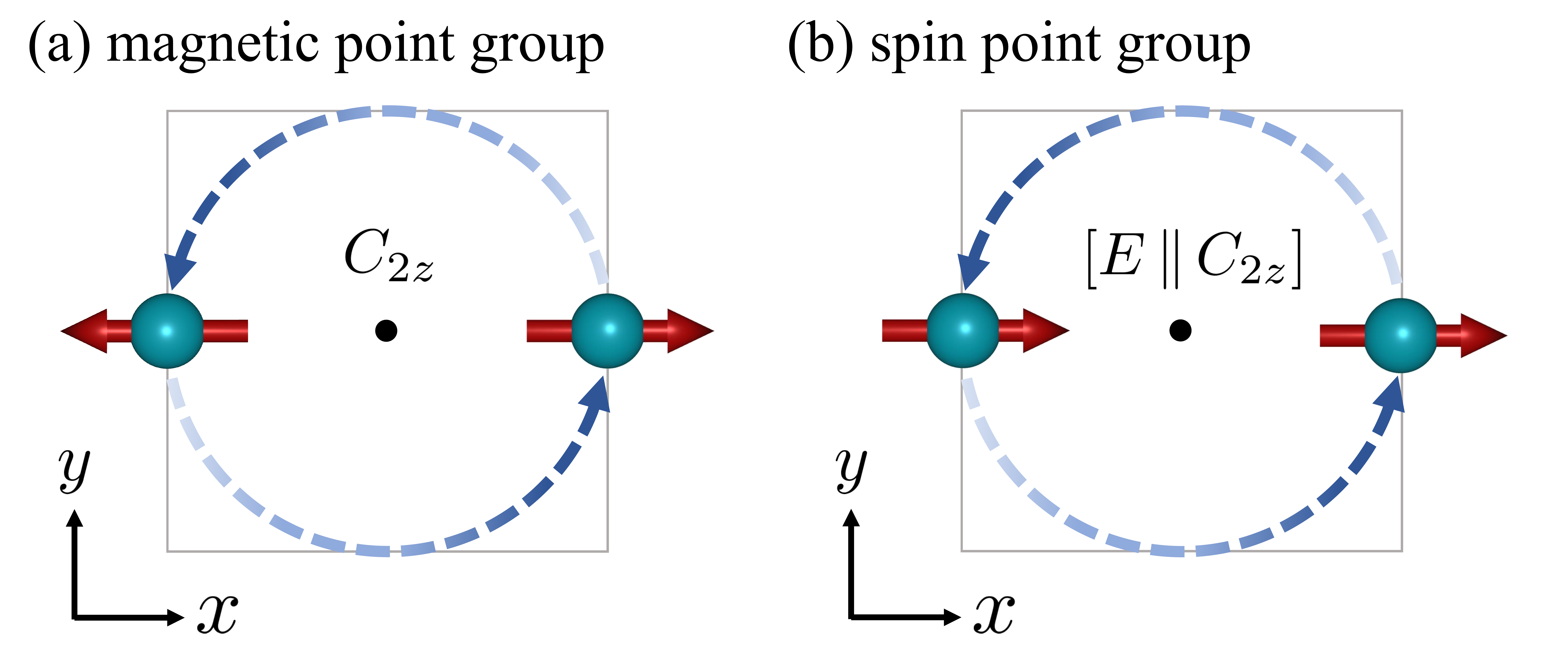}
    \caption{(Color online) Lattice-rotation operation 
    under (a) magnetic point group and (b) spin point group.
    In (a), the twofold rotation ($C_{2z}$) acts on both the orbital and spin degrees of freedom, whereas in (b), the operation ($[E \,\|\, C_{2z}]$) acts only on the orbital degree of freedom. 
    }
    \label{fig:rotation}
\end{figure}

The framework of multipoles offers symmetry-based perspectives for physical responses in solids~\cite{kuramoto2009multipole, Santini_RevModPhys.81.807, suzuki2018first, kusunose2022generalization, hayami2024unified}.
Many electric, magnetic, and transport responses can be interpreted as manifestations of active multipole degrees of freedom~\cite{Spaldin_0953-8984-20-43-434203, hayami2016emergent, Suzuki_PhysRevB.95.094406, Hayami_PhysRevB.98.165110, thole2018magnetoelectric, Watanabe_PhysRevB.98.245129, Hayami_PhysRevLett.122.147602, kimata2021x}. 
For instance, magnetic octupole $M_{\alpha\beta\gamma}$ becomes a microscopic order parameter of $\mathcal{T}$-broken antiferromagnets, including $d$-wave altermagnets~\cite{Bhowal_PhysRevX.14.011019, McClarty_PhysRevLett.132.176702, schiff2025collinear, sato2025quantum, oike2025thermodynamic}. 
When the SOC is present, the possible responses of a magnetic material can be understood by examining how the relevant multipoles are classified into the irreducible representations (IRREPs) of the magnetic point group (MPG)~\cite{Yatsushiro_PhysRevB.104.054412}. 
Meanwhile, when the effect of SOC is negligibly small, the applicability of this analysis becomes limited because MPGs inherently suppose inseparable coupling between orbital (lattice) and spin degrees of freedom, as shown in Fig.~\ref{fig:rotation}(a)~\cite{PhysRevLett.119.187204}.

To circumvent this limitation inherent to MPG, the concept of spin space group or spin point group (SPG), first introduced in the 1960s~\cite{Naish1962, Kitz1965, Brinkman1966,Litvin_Opechowski1974,Litvin1977}, has been developed into a more general symmetry framework applicable to SOC-free systems~\cite{Liu2022, Watanabe2024, Xiao2024, Chen2024,Jiang2024, Chen2025, SciPostPhys.18.3.109,Etxebarria2025}. 
In contrast to MPG, SPG consists of combined operations acting independently on 
orbital and spin space, allowing rotations in real space that are not accompanied by a corresponding rotation of the spin [Fig.~\ref{fig:rotation}(b)]. 
The overall spin point group $\mathcal{G}$ is expressed as the direct product 
of the spin-only group $\mathcal{G}_{\rm SO}$ and the nontrivial spin point group $\mathcal{G}_{\rm NT}$~\cite{Litvin_Opechowski1974}.
$\mathcal{G}_{\rm SO}$ contains purely spin-space operations, and has four types corresponding to nonmagnetic (paramagnetic), collinear, coplanar, and noncoplanar orders, while Litvin has enumerated a total of 598 nontrivial spin point groups $\mathcal{G}_{\rm NT}$~\cite{Litvin1977}, 
90 of which are compatible with collinear magnetism.
This framework captures SOC-free spin splittings, spin-current responses, and related effects, making SPG a natural basis for identifying SOC-free responses and their associated multipoles.

In the present study, we classify four types of multipoles, consisting of electric, magnetic, magnetic toroidal, and electric toroidal, that are activated in nonrelativistic collinear antiferromagnets based on SPG symmetry. 
By analyzing how multipoles transform under collinear SPG operations, we establish a correspondence between IRREPs and multipoles. 
Upon decomposing the multipoles into orbital and spin parts, we find that orbital-type electric and electric toroidal multipoles with $\mathcal{T}=+1$ and spin-type electric and electric toroidal multipoles with $\mathcal{T}=-1$ can belong to the identity IRREPs, while orbital- and spin-type magnetic and magnetic toroidal multipoles cannot. 
We also discuss the similarities and differences between SPGs and MPGs.
To illustrate the applicability of our framework, we present microscopic calculations of spin current generation by elastic waves free from SOC and magnetoelectric effect that relies on SOC for a specific magnetic structure.
Our results contribute to the exploration of nonrelativistic functional properties in nonrelativistic collinear antiferromagnets.

Let us first outline a systematic procedure for deriving SPGs associated with collinear magnetic order~\cite{Litvin1977, SciPostPhys.18.3.109}. 
In SPG, the operation $[S \,\|\, R]$ consists of a spin-space operation $S$ and orbital-space operation $R$.
For example, the $\mathcal{P}$ operation is represented as $[E \,\|\, I=\bar{1}=-\bm{1}]$ ($E$ is an identity operator) and the $\mathcal{T}$ operation as an ``inversion" in spin space $[\theta=1'=-\bm{1} \,\|\, E]$, though alternative conventions are sometimes used~\cite{Etxebarria2025}. 
It is noted that $[\theta \,\|\, E]$ also represents the sign reversal of momentum.

For collinear antiferromagnetic structures with magnetic moments oriented along $\bm{n}$, spin-only group is expressed as a semidirect product 
$\mathcal{G}_{\rm SO}^\infty  =$ SO(2) $\rtimes \{ [E \,\|\, E]$, $[\theta 2_{\perp\bm{n}} \,\|\, E] \}$, where 
$2_{\perp\bm{n}}$ denotes a $\pi$-rotation perpendicular to $\bm{n}$ 
and SO(2) represents the continuous group from the arbitrariness of $2_{\perp\bm{n}}$. 
Hereafter, $\bm{n}$ is fixed at $(0,0,1)$ without loss of generality.

There are 90 nontrivial spin point groups compatible with $\mathcal{G}_{\rm SO}^\infty$: 32 unitary and 58 non-unitary groups. 
Let $\bm{G}$ be the spatial parent point group 
(one of the 32 crystallographic point groups), and let $\bm{H}$ 
denote a halving unitary subgroup of $\bm{G}$. 
Then, the nontrivial spin point group $\mathcal{G}_{\rm NT}$ is constructed as follows:
\begin{align}
\mathcal{G}_{\rm NT} 
=
\begin{cases}
\bm{G} ,\quad  \\
\bm{H} + [\theta  \,\|\, E] (\bm{G} - \bm{H}) ,
\end{cases}
\end{align}
where the upper and lower correspond to unitary and non-unitary spin point groups, respectively. 
The operations included in $\bm{G}$ and $\bm{H}$ 
act only in real space. 
With an operator $[E \,\|\, R_0]$ chosen from $\bm{G} - \bm{H}$, the non-unitary spin point group is rewritten as $\bm{H} + [\theta \,\|\, R_0]\bm{H}$, which means that the nonunitary part is generated by combining the antiunitary operation $[\theta \,\|\, R_0]$ with all elements of the unitary subgroup $\bm{H}$.
In the end, the overall collinear spin point group $\mathcal{G}$ is written as 
\begin{align}
\mathcal{G}
&=
\mathcal{G}_{\rm SO}^\infty \times \mathcal{G}_{\rm NT} 
\\
&=
\mathrm{SO(2)} \times
\begin{cases}
\bm{G} + [\theta 2_{\perp\bm{n}} \,\|\, E] \bm{G} , \quad  \\
\bm{H} + [\theta  \,\|\, R_0] \bm{H} \\
\quad + [\theta 2_{\perp\bm{n}} \,\|\, E](\bm{H} + [\theta  \,\|\, R_0] \bm{H})  .
\end{cases}
\end{align}
The IRREPs of $\mathcal{G}$ are obtained through the IRREPs of $\mathcal{G}_{\rm NT}$. 
Although deriving the IRREPs is trivial when $\mathcal{G}_{\rm NT}$ is unitary, the nonunitary case requires a slightly different procedure. 
Once the halving subgroup $\bm{H}$ is identified, the IRREPs of $\mathcal{G}$ can be obtained by augmenting the IRREPs of $\bm{H}$ with appropriate 
indices for the antiunitary operators $[\theta  \,\|\, R_0]$ and $[\theta 2_{\perp\bm{n}} \,\|\, E]$.
We attach superscripts corresponding to $[\theta  \,\|\, R_0]$ to the IRREPs of $\bm{H}$, and also add a bar over them when the parity under 
$[\theta 2_{\perp\bm{n}} \,\|\, E]$ is $-1$.

\renewcommand{\arraystretch}{1.5} % 表の縦幅: 1.0 → 1.5 (可能であれば) 
\begin{table}[h] 
\begin{center}
\caption{
\label{Parities}
Classification of the four types of multipoles (Electric (E): $Q_{lm}$, Magnetic (M): $M_{lm}$, Magnetic toroidal (MT): $T_{lm}$, Electric toroidal (ET): $G_{lm}$) according to their spatial inversion $(\mathcal{P})$, time-reversal $(\mathcal{T})$, and combined ($\mathcal{PT}$) parities, with those activated in collinear spin point groups (CSPG) indicated by checkmarks $(\checkmark)$. 
Panels (a) and (b) list the orbital and spin multipoles, respectively. 
}
(a) orbital multipoles $X^{\rm (o)}$
\begin{tabular}{cccccc} \hline\hline
Type
& Notation
& $\mathcal{P}$ & \, \,  $\mathcal{T}$ \, \,  & $\mathcal{PT}$ & CSPG 
\\ \hline 
E
& $Q_{lm}^{\rm (o)}$
& $(-1)^l$ & $+1$ & $(-1)^l$ & $\checkmark$  
\\ 
M
& $M_{lm}^{\rm (o)}$
& $(-1)^{l+1}$ & $-1$ & $(-1)^l$ &    
\\  
MT
& $T_{lm}^{\rm (o)}$
& $(-1)^l$ & $-1$ & $(-1)^{l+1}$ &  
\\ 
ET
& $G_{lm}^{\rm (o)}$
& $(-1)^{l+1}$ & $+1$ & $(-1)^{l+1}$ & $\checkmark$ 
\\ \hline\hline 
\end{tabular}
(b) spin multipoles $X^{\rm (s)} \equiv \sigma_z X^{\rm (o)}$ 
\begin{tabular}{cccccc} \hline\hline
Type  
& Notation
& $\mathcal{P}$ & \, \,  $\mathcal{T}$ \, \,  & $\mathcal{PT}$ & CSPG
\\ \hline  
E
& $Q_{lm}^{\rm (s)}$
& $(-1)^l$ & $-1$ & $(-1)^{l+1}$ & $\checkmark$ 
\\ 
M 
& $M_{lm}^{\rm (s)}$
& $(-1)^{l+1}$ & $+1$ & $(-1)^{l+1}$ &   
\\  
MT 
& $T_{lm}^{\rm (s)}$
& $(-1)^l$ & $+1$ & $(-1)^l$ & 
\\  
ET
& $G_{lm}^{\rm (s)}$
& $(-1)^{l+1}$ & $-1$ & $(-1)^l$ & $\checkmark$ 
\\ \hline\hline
\end{tabular}
\end{center}
\end{table}

\begin{table*}[htb!]
\caption{
IRREPs of multipoles $(l\leq 4)$ on the collinear spin point group 
$\mathcal{G}_{\rm SO}^\infty  \times {}^{\bar{1}} 4/ {}^{\bar{1}} m {}^1 m {}^{\bar{1}} m 
= \mathcal{G}_{\rm SO}^\infty  \times (\bar{4}m2 + [\theta \,\|\, I] \ \bar{4}m2)$. 
In the IRREPs, the sign $\pm$ represents the parity with respect to the antiunitary operation $[\theta \,\|\, I]$, while 
the absence [presence] of a bar over the IRREPs indicates parity $+1$ [$-1$] under the antiunitary operation $[\theta 2_{\perp\bm{n}} \,\|\, E]$.
Primary and secondary axes with respect to the symmetry operations in the Cartesian coordinates are [001] and [110], respectively.
}
\label{IRREPs}
\centering
\begingroup
\renewcommand{\arraystretch}{1.4} % 表の縦幅 1.0 → 1.4 (可能であれば) 
\begin{tabular}{cc|cc}
\hline\hline
IRREP 
& E, ET & 
IRREP
& M, MT \\
\hline
${\rm A_1^+}$
& $Q_{0}^{\rm (o)}$, $Q_{u}^{\rm (o)}$, $Q_{4}^{\rm (o)}$, $Q_{4u}^{\rm (o)}$, $Q_{z}^{\beta {\rm (s) } }$, 
$G_{xy}^{\rm (s)}$, $G_{4z}^{\beta {\rm (s) } }$ 
&
${\rm \bar{A}_1^+}$
& $M_{xy}^{\rm (o)}$, $M_{4z}^{\beta {\rm (o) } }$,
$T_{z}^{\beta {\rm (o) } }$, 
$T_{0}^{\rm (s)}$, $T_{u}^{\rm (s)}$, $T_{4}^{\rm (s)}$, $T_{4u}^{\rm (s)}$ 
\\ 
${\rm A_1^-}$
& $Q_{z}^{\beta {\rm (o) } }$, 
$Q_{0}^{\rm (s)}$, $Q_{u}^{\rm (s)}$, $Q_{4}^{\rm (s)}$, $Q_{4u}^{\rm (s)}$,
$G_{xy}^{\rm (o)}$, $G_{4z}^{\beta {\rm (o) } }$ 
&
${\rm \bar{A}_1^-}$
& $M_{xy}^{\rm (s)}$, $M_{4z}^{\beta {\rm (s) } }$,
$T_{0}^{\rm (o)}$, $T_{u}^{\rm (o)}$, $T_{4}^{\rm (o)}$, $T_{4u}^{\rm (o)}$, $T_{z}^{\beta {\rm (s) } }$
\\ 
\hline
${\rm A_2^+}$
& $Q_{4z}^{\alpha {\rm (o) } }$,
$Q_{xyz}^{\rm (s)}$, 
$G_{z}^{\rm (o)}$, $G_{z}^{\alpha {\rm (o) } }$,
$G_{v}^{\rm (s)}$, $G_{4v}^{\rm (s)}$
&
${\rm \bar{A}_2^+}$
& $M_{v}^{\rm (o)}$, $M_{4v}^{\rm (o)}$,
$M_{z}^{\rm (s)}$, $M_{z}^{\alpha {\rm (s) } }$, 
$T_{xyz}^{\rm (o)}$,
$T_{4z}^{\alpha {\rm (s) } }$
\\ 
${\rm A_2^-}$ 
& $Q_{xyz}^{\rm (o)}$,
$Q_{4z}^{\alpha {\rm (s) } }$,
$G_{v}^{\rm (o)}$, $G_{4v}^{\rm (o)}$,
$G_{z}^{\rm (s)}$, $G_{z}^{\alpha {\rm (s) } }$
&
${\rm \bar{A}_2^-}$
& $M_{z}^{\rm (o)}$, $M_{z}^{\alpha {\rm (o) } }$,
$M_{v}^{\rm (s)}$, $M_{4v}^{\rm (s)}$,
$T_{4z}^{\alpha {\rm (o) } }$,
$T_{xyz}^{\rm (s)}$
\\ 
\hline
${\rm B_1^+}$
& $Q_{xy}^{\rm (o)}$, $Q_{4z}^{\beta {\rm (o) } }$,
$G_{z}^{\beta {\rm (o) } }$,
$G_{0}^{\rm (s)}$, $G_{u}^{\rm (s)}$, $G_{4}^{\rm (s)}$, $G_{4u}^{\rm (s)}$
&
${\rm \bar{B}_1^+}$
& $M_{0}^{\rm (o)}$, $M_{u}^{\rm (o)}$, $M_{4}^{\rm (o)}$, $M_{4u}^{\rm (o)}$,
$M_{z}^{\beta {\rm (s) } }$,
$T_{xy}^{\rm (s)}$, $T_{4z}^{\beta {\rm (s) } }$ 
\\ 
${\rm B_1^-}$
& $Q_{xy}^{\rm (s)}$, $Q_{4z}^{\beta {\rm (s) } }$,
$G_{0}^{\rm (o)}$, $G_{u}^{\rm (o)}$, $G_{4}^{\rm (o)}$, $G_{4u}^{\rm (o)}$,
$G_{z}^{\beta {\rm (s) } }$
&
${\rm \bar{B}_1^-}$ 
& $M_{z}^{\beta {\rm (o) } }$,
$M_{0}^{\rm (s)}$, $M_{u}^{\rm (s)}$, $M_{4}^{\rm (s)}$, $M_{4u}^{\rm (s)}$,
$T_{xy}^{\rm (o)}$, $T_{4z}^{\beta {\rm (o) } }$ 
\\
\hline
${\rm B_2^+}$ 
& $Q_{v}^{\rm (o)}$, $Q_{4v}^{\rm (o)}$,
$Q_{z}^{\rm (s)}$, $Q_{z}^{\alpha {\rm (s) } }$,
$G_{xyz}^{\rm (o)}$,
$G_{4z}^{\alpha {\rm (s) } }$
&
${\rm \bar{B}_2^+}$
& $M_{4z}^{\alpha {\rm (o) } }$,
$M_{xyz}^{\rm (s)}$,
$T_{z}^{\rm (o)}$, $T_{z}^{\alpha {\rm (o) } }$,
$T_{v}^{\rm (s)}$, $T_{4v}^{\rm (s)}$
\\ 
${\rm B_2^-}$ 
& $Q_{z}^{\rm (o)}$, $Q_{z}^{\alpha {\rm (o) } }$,
$Q_{v}^{\rm (s)}$, $Q_{4v}^{\rm (s)}$,
$G_{4z}^{\alpha {\rm (o) } }$,
$G_{xyz}^{\rm (s)}$
&
${\rm \bar{B}_2^-}$ 
& $M_{xyz}^{\rm (o)}$,
$M_{4z}^{\alpha {\rm (s) } }$,
$T_{v}^{\rm (o)}$, $T_{4v}^{\rm (o)}$,
$T_{z}^{\rm (s)}$, $T_{z}^{\alpha {\rm (s) } }$ 
\\ 
\hline
${\rm E^+}$
& \{$Q_{yz}^{\rm (o)}$, $Q_{zx}^{\rm (o)}$\}, \{$Q_{4x}^{\alpha {\rm (o) } }$, $Q_{4y}^{\alpha {\rm (o) } }$\}, \{$Q_{4x}^{\beta {\rm (o) } }$, $Q_{4y}^{\beta {\rm (o) } }$\},
&
${\rm \bar{E}^+}$
& \{$M_{yz}^{\rm (o)}$, $M_{zx}^{\rm (o)}$\}, \{$M_{4x}^{\alpha {\rm (o) } }$, $M_{4y}^{\alpha {\rm (o) } }$\}, \{$M_{4x}^{\beta {\rm (o) } }$, $M_{4y}^{\beta {\rm (o) } }$\},
\\ 
& \{$Q_{x}^{\rm (s)}$, $Q_{y}^{\rm (s)}$\},  \{$Q_{x}^{\alpha {\rm (s) } }$, $Q_{y}^{\alpha {\rm (s) } }$\}, \{$Q_{x}^{\beta {\rm (s) } }$, $Q_{y}^{\beta {\rm (s) } }$\},
&
& \{$M_{x}^{\rm (s)}$, $M_{y}^{\rm (s)}$\}, \{$M_{x}^{\alpha {\rm (s) } }$, $M_{y}^{\alpha {\rm (s) } }$\}, \{$M_{x}^{\beta {\rm (s) } }$, $M_{y}^{\beta {\rm (s) } }$\},
\\
& \{$G_{x}^{\rm (o)}$, $G_{y}^{\rm (o)}$\}, \{$G_{x}^{\alpha {\rm (o) } }$, $G_{y}^{\alpha {\rm (o) } }$\}, \{$G_{x}^{\beta {\rm (o) } }$, $G_{y}^{\beta {\rm (o) } }$\},
&
& \{$T_{x}^{\rm (o)}$, $T_{y}^{\rm (o)}$\}, \{$T_{x}^{\alpha {\rm (o) } }$, $T_{y}^{\alpha {\rm (o) } }$\}, \{$T_{x}^{\beta {\rm (o) } }$, $T_{y}^{\beta {\rm (o) } }$\},
\\ 
& \{$G_{yz}^{\rm (s)}$, $G_{zx}^{\rm (s)}$\}, \{$G_{4x}^{\alpha {\rm (s) } }$, $G_{4y}^{\alpha {\rm (s) } }$\}, \{$G_{4x}^{\beta {\rm (s) } }$, $G_{4y}^{\beta {\rm (s) } }$\} 
& 
& \{$T_{yz}^{\rm (s)}$, $T_{zx}^{\rm (s)}$\}, \{$T_{4x}^{\alpha {\rm (s) } }$, $T_{4y}^{\alpha {\rm (s) } }$\}, \{$T_{4x}^{\beta {\rm (s) } }$, $T_{4y}^{\beta {\rm (s) } }$\}
\\ 
${\rm E^-}$
& \{$Q_{x}^{\rm (o)}$, $Q_{y}^{\rm (o)}$\},  \{$Q_{x}^{\alpha {\rm (o) } }$, $Q_{y}^{\alpha {\rm (o) } }$\}, \{$Q_{x}^{\beta {\rm (o) } }$, $Q_{y}^{\beta {\rm (o) } }$\},
&
${\rm \bar{E}^-}$
& \{$M_{x}^{\rm (o)}$, $M_{y}^{\rm (o)}$\}, \{$M_{x}^{\alpha {\rm (o) } }$, $M_{y}^{\alpha {\rm (o) } }$\}, \{$M_{x}^{\beta {\rm (o) } }$, $M_{y}^{\beta {\rm (o) } }$\},
\\
& \{$Q_{yz}^{\rm (s)}$, $Q_{zx}^{\rm (s)}$\}, \{$Q_{4x}^{\alpha {\rm (s) } }$, $Q_{4y}^{\alpha {\rm (s) } }$\}, \{$Q_{4x}^{\beta {\rm (s) } }$, $Q_{4y}^{\beta {\rm (s) } }$\},
&
& \{$M_{yz}^{\rm (s)}$, $M_{zx}^{\rm (s)}$\}, \{$M_{4x}^{\alpha {\rm (s) } }$, $M_{4y}^{\alpha {\rm (s) } }$\}, \{$M_{4x}^{\beta {\rm (s) } }$, $M_{4y}^{\beta {\rm (s) } }$\},
\\
& \{$G_{yz}^{\rm (o)}$, $G_{zx}^{\rm (o)}$\}, \{$G_{4x}^{\alpha {\rm (o) } }$, $G_{4y}^{\alpha {\rm (o) } }$\}, \{$G_{4x}^{\beta {\rm (o) } }$, $G_{4y}^{\beta {\rm (o) } }$\}, 
&
& \{$T_{yz}^{\rm (o)}$, $T_{zx}^{\rm (o)}$\}, \{$T_{4x}^{\alpha {\rm (o) } }$, $T_{4y}^{\alpha {\rm (o) } }$\}, \{$T_{4x}^{\beta {\rm (o) } }$, $T_{4y}^{\beta {\rm (o) } }$\},
\\
& \{$G_{x}^{\rm (s)}$, $G_{y}^{\rm (s)}$\}, \{$G_{x}^{\alpha {\rm (s) } }$, $G_{y}^{\alpha {\rm (s) } }$\}, \{$G_{x}^{\beta {\rm (s) } }$, $G_{y}^{\beta {\rm (s) } }$\}
&
& \{$T_{x}^{\rm (s)}$, $T_{y}^{\rm (s)}$\}, \{$T_{x}^{\alpha {\rm (s) } }$, $T_{y}^{\alpha {\rm (s) } }$\}, \{$T_{x}^{\beta {\rm (s) } }$, $T_{y}^{\beta {\rm (s) } }$\}
\\
\hline\hline
\end{tabular}
\endgroup
\end{table*}

We subsequently carry out the multipole classification for the collinear SPGs obtained in this manner. 
Multipoles consist of electric, magnetic, magnetic toroidal, and electric toroidal multipoles with distinct $\mathcal{P}$ and $\mathcal{T}$ parities, which are denoted as $Q_{lm}$, $M_{lm}$, $T_{lm}$, and $G_{lm}$, respectively~\cite{hayami2024unified}; the subscripts $l$ and $m$ denote orbital angular momentum (rank of multipoles) and its component. 
Although such multipoles are usually defined in a form that mixes spin and orbital degrees of freedom~\cite{kusunose2020complete}, the separation of spin and orbital symmetry operations in spin groups makes it convenient to redefine them independently. 
To this end, we distinguish between the orbital and spin components for each multipole $X$; we refer to ``orbital multipole $X^{\rm (o)}$" in the spinless space and ``spin multipole $X^{\rm (s)}  \equiv \sigma_z X^{\rm (o)}$" in the spinful space; $\sigma_z$ denotes the $z$ component of the Pauli matrix in spin space. 
It is noted that considering a single spin component is sufficient for describing collinear antiferromagnets. 
The parities of $X^{\rm (o)}$ and $X^{\rm (s)}$ in terms of $\mathcal{P}$, $\mathcal{T}$, and $\mathcal{PT}$ are summarized in Table~\ref{Parities}.

From the result in Table~\ref{Parities}, one finds that, among orbital multipoles, only the electric and electric toroidal types with $\mathcal{T}=+1$ can be active, whereas, among spin multipoles, the electric and electric toroidal types with $\mathcal{T}=-1$ can be active, since electric and electric toroidal multipoles are always even under $[\theta 2_{\perp\bm{n}} \,\|\, E]$ in collinear SPGs. 
This difference in the time-reversal parity $\mathcal{T}$ between orbital and spin multipoles originates from their different transformation properties under the operation $[\theta 2_{\perp\bm{n}} \,\|\, E]$. 
For orbital multipoles, which do not carry spin degrees of freedom, this operation effectively reduces to the time-reversal operation $[\theta  \,\|\, E]$, allowing only $\mathcal{T}=+1$ components, whereas spin multipoles are constructed as $X^{\rm (s)} = \sigma_z X^{\rm (o)}$, where $\sigma_z$ is odd under time reversal but even under $[\theta 2_{\perp\bm{n}} \,\|\, E]$, thereby permitting $\mathcal{T}=-1$ components. 
In contrast, magnetic and magnetic toroidal multipoles do not appear as active multipoles in collinear SPGs because their parities under the operation $[\theta 2_{\perp\bm{n}} \,\|\, E]$ are always $-1$. 
This indicates that orbital quantities 
represented by $M_{lm}^{\rm (o)}$ and $T_{lm}^{\rm (o)}$, such as orbital magnetization, 
%, such as orbital currents, 
cannot be activated without SOC in collinear magnetic systems, consistent with the \lq\lq orbital time-reversal symmetry" described in Ref.~\citen{Watanabe2024}. 
These results hold for all collinear SPGs.

\begin{figure}[t]
    \centering
    \includegraphics[width=1.0\linewidth]{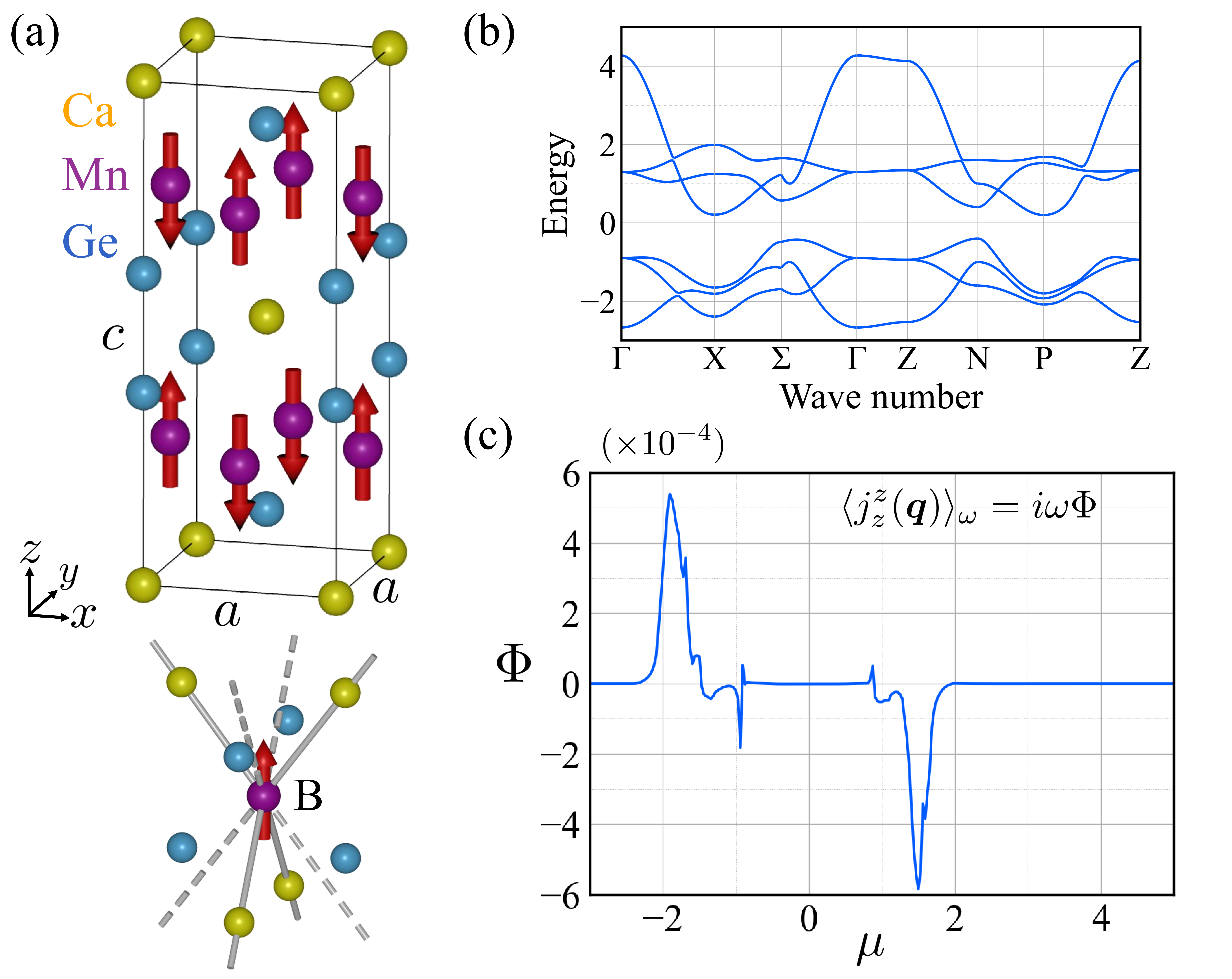}
    \caption{(Color online) (a) (Top panel) 
    Crystal and magnetic structures of CaMn$_2$Ge$_2$. 
    The spin point group of the system is 
    $\mathcal{G}_{\rm SO}^\infty \times {}^{\bar{1}} 4/ {}^{\bar{1}} m {}^1 m {}^{\bar{1}} m$ 
    and its magnetic point group is $4'/m'm'm$. 
    In the bottom panel, solid and dashed lines represent the inequivalent hopping strengths between the A (down spin moments) and B (up spin moments) sublattices over the distance $\sqrt{a^2 + (c/2)^2}$. 
    (b) Band structure along the high-symmetry lines, where 
    $\Gamma$: $\bm{0}$, 
    ${\rm X}$: $\bm{b}_3/2$, 
    $\Sigma$: $a^2/(2c^2)(-\bm{b}_1+\bm{b}_2+\bm{b}_3)$, 
    ${\rm Z}$: $(\bm{b}_1+\bm{b}_2-\bm{b}_3)/2$, 
    ${\rm N}$: $\bm{b}_2/2$, 
    and ${\rm P}$: $(\bm{b}_1+\bm{b}_2+\bm{b}_3)/4$ 
    with the primitive reciprocal lattice vectors 
    $\bm{b}_1=(2\pi/a)\hat{\bm{y}}+(2\pi/c)\hat{\bm{z}}$, 
    $\bm{b}_2=(2\pi/a)\hat{\bm{x}}+(2\pi/c)\hat{\bm{z}}$, and 
    $\bm{b}_3=(2\pi/a)\hat{\bm{x}}+(2\pi/a)\hat{\bm{y}}$.
    (c) Chemical potential dependence of the spin current induced by the longitudinal elastic waves. 
    }
    \label{fig:jzz}
\end{figure}

Let us consider an example of the multipole classification for the specific collinear magnetic structure under tetragonal symmetry, as shown in Fig.~\ref{fig:jzz}(a); such a magnetic structure is found in materials like CaMn$_2$Ge$_2$~\cite{Malaman1994, Di_Napoli2007}.
The symmetry groups of this system are given by $\bm{G}=4/mmm$, $\bm{H}=\bar{4}m2$, 
and  
$\mathcal{G}_{\rm NT} = {}^{\bar{1}} 4/ {}^{\bar{1}} m {}^1 m {}^{\bar{1}} m$ (No.~189) 
according to Litvin's notation~\cite{Litvin1977} 
and character table for the collinear spin point group $\mathcal{G}_{\rm SO}^\infty \times {}^{\bar{1}} 4/ {}^{\bar{1}} m {}^1 m {}^{\bar{1}} m$ is enumerated in Supplemental Materials (SM)~\cite{suppl}. 
The classification result is summarized in Table~\ref{IRREPs}, where we use the multipole notations in Refs.~\citen{Hayami_PhysRevB.98.165110, Yatsushiro_PhysRevB.104.054412}; the rank-0 monopole is denoted as $X_0$, the rank-1 dipole as ($X_x$, $X_y$, $X_z$), the rank-2 quadrupole as 
($X_u$, $X_v$, $X_{yz}$, $X_{zx}$, $X_{xy}$), 
the rank-3 octupole as 
($X_{xyz}$, $X_x^\alpha$, $X_y^\alpha$, $X_z^\alpha$, $X_x^\beta$, $X_y^\beta$, $X_z^\beta$), and the rank-4 hexadecapole as 
($X_4$, $X_{4u}$, $X_{4v}$, $X_{4x}^\alpha$, $X_{4y}^\alpha$, $X_{4z}^\alpha$, $X_{4x}^\beta$, $X_{4y}^\beta$, $X_{4z}^\beta$).
Among the active multipoles belonging to the identity IRREP, A$_1^+$, the spin electric octupole 
($Q_{z}^{\beta {\rm (s) } }$) and spin electric toroidal multipoles 
($G_{xy}^{\rm (s)}$, $G_{4z}^{\beta {\rm (s) } }$) are activated by collinear antiferromagnetic ordering in Fig.~\ref{fig:jzz}(a), corresponding to the order parameter, whereas the others remain in the high-temperature paramagnetic state.

Within the multipoles belonging to the identity IRREP, we focus on 
$Q_{z}^{\beta {\rm (s) } }$, whose functional form is given by 
$(x^2 -y^2)z\sigma_z$. 
This fourth-order coupling indicates the emergence of the SOC-free spin-current response against the strain; the spin current $j_z^z \propto j_z\sigma_z$ 
is generated by the $(u_{xx}-u_{yy})$-type strain, since the symmetry of $j_z^z (u_{xx}-u_{yy})$ is the same as $(x^2 -y^2)z\sigma_z$ except for the time-reversal parity.
The difference in time-reversal parity means that the above response is AC-type (time-derivative) or dissipative. 
We demonstrate such cross-correlation responses in the following model analysis.

The expected physical responses are qualitatively different once the SOC is taken into account. 
The entanglement of orbital and spin operations in the presence of the SOC removes the distinction between barred and unbarred IRREPs, reducing to the MPG.
In the case of $\mathcal{G}_{\rm SO}^\infty  \times {}^{\bar{1}} 4/ {}^{\bar{1}} m {}^1 m {}^{\bar{1}} m$, 
two IRREPs ${\rm A_1^+}$ and ${\rm \bar{A}_2^+}$ belong to the identity IRREP of the MPG $4'/m'm'm$, since the SPG operations $\{ [E \,\|\, E]$, $[\theta 2_{\perp\bm{n}}  \,\|\, E] \}\times$$\{ [ E \,\|\, C_{2[110]} ]$, $[ E \,\|\, C_{2[\bar{1}10]} ]$, $[ \theta  \,\|\, C_{2[100]} ]$, $[ \theta \,\|\, C_{2[010]} ] \}$ 
reduce to $\{ C_{2[100]}$, $C_{2[010]}$, $\theta C_{2[110]}$, $\theta C_{2[\bar{1}10]} \}$ under nonzero SOC. 
Thus, the orbital magnetic quadrupole $M_v^{\rm (o)}$ and other odd-parity magnetic and magnetic toroidal multipoles belonging to ${\rm \bar{A}_2^+}$ can be induced by the SOC. 
Accordingly, the linear magnetoelectric effect, where net magnetization is induced by an external electric field, is expected~\cite{dzyaloshinskii1960magneto, astrov1960magnetoelectric,Khanh_PhysRevB.93.075117,Yanagi_PhysRevB.97.020404}. 
We discuss such a response under the SOC in SM~\cite{suppl}.

To demonstrate spin current generation by elastic waves under the above magnetic structure without the SOC, we analyze a tight-binding Hamiltonian of CaMn$_2$Ge$_2$ by incorporating three $d$ orbitals on Mn ions in Fig.~\ref{fig:jzz}(a), which is given by 
$H_0 = H_{\rm hop} + H_{\rm MF}$,
with
\begin{align}
 H_{\rm hop} 
 &= 
 \displaystyle \sum_{\bm{r}_i,\bm{r}_j} \sum_{\alpha,\beta} \sum_{\mu, \nu } 
    E_{\mu,\nu}(\bm{r}_{j\beta}-\bm{r}_{i\alpha}) c_{i,\alpha,\mu}^\dagger c_{j,\beta,\nu} ,  
\\
 H_{\rm MF} 
 &= 
 h \displaystyle \sum_{\bm{r}_i} \sum_{\mu}
    ( \, 
    c_{i,\mathrm{A},\mu}^\dagger \, \sigma_z \, c_{i,\mathrm{A},\mu} 
    - c_{i,\mathrm{B},\mu}^\dagger \, \sigma_z \, c_{i,\mathrm{B},\mu} 
    ) ,
\end{align}
where a two-spin-component operator, 
$c_{i,\alpha,\mu} =$ ${}^t (c_{i,\alpha,\mu, \uparrow}$, $c_{i,\alpha, \mu, \downarrow})$,  
that annihilates an electron in orbital $\mu$ ($= d_{xy}, d_{yz}, d_{zx}$) at sublattice $\alpha$ ($=$ A, B) in the $i$-th unit cell is introduced.
$H_{\rm hop}$ represents the hopping Hamiltonian, where $E_{\mu,\nu}(\bm{r}_{j\beta}-\bm{r}_{i\alpha})$ is the orbital-dependent hopping integral between the lattice sites at $\bm{r}_{i\alpha}$ and $\bm{r}_{j\beta}$ 
($\bm{r}_{i\alpha}=\bm{r}_i + \bm{\xi}_\alpha$ 
is the coordinate of the sublattice $\alpha$ in the $i$-th unit cell at $\bm{r}_i$).
Angular dependence of $E_{\mu,\nu}$ is given by Slater-Koster parameters~\cite{Slater1954}, 
$t_\sigma$, $t_\pi$, and $t_\delta$.
We consider the 1st, 2nd, and 3rd nearest-neighbor hoppings ($\abs{\bm{r}_{j\beta}-\bm{r}_{i\alpha}} = a/\sqrt{2}$, $a$, $c/2$), and we also introduce hoppings at the distance of $\sqrt{a^2 + (c/2)^2}$ with two different strengths by the factor 1.5 depending on the bond direction, as shown in the bottom panel of Fig.~\ref{fig:jzz}(a). 
This contribution is necessary to reflect the correct SPG symmetry for the underlying magnetic structure in CaMn$_2$Ge$_2$; see SM~\cite{suppl} for the details. 
The hopping parameters are given $(t_\sigma, t_\pi, t_\delta)=(1, -0.5, 0.25)$ for $\abs{\bm{r}_{j\beta}-\bm{r}_{i\alpha}} = a/\sqrt{2}$, 
$(t'_\sigma, t'_\pi, t'_\delta)=(0.4, -0.2, 0.1 )$ for $\abs{\bm{r}_{j\beta}-\bm{r}_{i\alpha}} = a$, 
$(t''_\sigma, t''_\pi, t''_\delta)=(0.2, -0.1, 0.05)$ for $\abs{\bm{r}_{j\beta}-\bm{r}_{i\alpha}} = c/2$,
and 
$(t'''_\sigma, t'''_\pi, t'''_\delta)=(0.1, -0.05, 0.025)$ for $\abs{\bm{r}_{j\beta}-\bm{r}_{i\alpha}} = \sqrt{a^2 + (c/2)^2}$.
We set $t_\sigma$ as the energy unit and lattice constant $a=c/2.6=1$ as the length unit. 
The second term, $H_{\mathrm{MF}}$, describes the mean field to induce the collinear magnetic structure in Fig.~\ref{fig:jzz}(a); we set $h=1$. 
Figure~\ref{fig:jzz}(b) shows the electronic band structure, where each band is doubly degenerate owing to the $\mathcal{PT}$ symmetry.

As the effect of the elastic wave, we consider the modulation of the 1st, 2nd, and 3rd nearest-neighbor hopping integrals~\cite{Ogawa2023}.
Using the angular dependence of $E_{\mu,\nu}$ and assuming that the hopping parameters are inversely proportional to the square of the distance, the perturbation Hamiltonian is obtained as
$\delta H = \sum_{\bm{k}} c_{\bm{k}+\bm{q}/2}^\dagger 
\delta \mathcal{H}(\bm{k};\bm{q})
c_{\bm{k}-\bm{q}/2} e^{-i\omega t}$; see SM~\cite{suppl} for the derivation.

We evaluate the $z$-polarized spin current flowing along the $z$-direction induced by elastic waves within the linear response theory.  
The spin current operator is defined as 
$j_z^{z} (\bm{q}) = \sum_{\bm{k}} 
c_{\bm{k}-\bm{q}/2}^\dagger
\, j_z^{z} \, 
c_{\bm{k}+\bm{q}/2}$ 
with 
$j_{z}^{z}\equiv 
\{  \partial \mathcal{H}_0(\bm{k})/\partial k_z, S_z \}_+ /2 $, 
where $H_0 = \sum_{\bm{k}} c_{\bm{k}}^\dagger \mathcal{H}_0 (\bm{k}) \, c_{\bm{k}}$ 
and $S_z$ is a matrix that represents total $z$-polarized spin operators of electrons.
We focus on the $\omega$-linear (and also $\bm{q}$-linear) term of the spin current 
$\langle{j_z^{z}}(\bm{q})\rangle_\omega = i\omega \Phi 
= i\omega  \lim_{\omega\rightarrow 0} \{ \chi(\bm{q}, \omega) - \chi(\bm{q}, 0) \}/(i \omega)$ [$\chi(\bm{q}, \omega)$ is the complex susceptibility], which is given by 
\begin{align}
    \Phi
    &=
    -\frac{1}{N}\sum_{\bm{k}} \sum_{m,n} 
    \frac{f(\varepsilon_{m\bm{k}}) - f(\varepsilon_{n\bm{k}})}{\varepsilon_{m\bm{k}} - \varepsilon_{n\bm{k}}}
    \nonumber
    \\
    &\quad\quad\quad
    \times 
    \Im{
    \frac{\bra{m\bm{k}} j_{z}^{z} \ket{n\bm{k}} \bra{n\bm{k}} \delta \mathcal{H}(\bm{k};\bm{q}) \ket{m\bm{k}}}{\varepsilon_{m\bm{k}} - \varepsilon_{n\bm{k}} + i\delta}
    },
\end{align}
where $f(E) = [ e^{(E - \mu)/k_{\rm B}T}+ 1]^{-1}$ 
is the Fermi--Dirac distribution function, with the temperature set to $k_{\rm B}T=10^{-3}$. 
$\varepsilon_{n\bm{k}}$ and $\ket{n\bm{k}}$ denote the eigenvalue and eigenstate of the 
Hamiltonian $\mathcal{H}_0(\bm{k})$ for band index $n$ and wave vector $\bm{k}$, respectively. 
$\delta$ represents the phenomenological broadening factor corresponding to the inverse of the relaxation time. 
The obtained spin current is independent of $\delta$ (in the regime of $\delta \ll 1$); we set $\delta = 10^{-3}$.
$N$ represents the number of unit cells, and we set $\hbar = 1$.
$\Phi$ is a dimensionless quantity that characterizes the magnitude of the spin current induced by elastic waves.

Figure~\ref{fig:jzz}(c) shows the chemical potential $\mu$ dependence of the induced spin current for $u_{xx}-u_{yy} = 0.01$. 
As expected from the symmetry, the response function $\Phi$ becomes nonzero even without the SOC. 
The enhancement of $\Phi$ at $\mu\approx -1.8$ and $1.5$ arises from the interband effect, where the Fermi level lies in regions of small band separation.
These results confirm that the spin current generation in this system is an intrinsically nonrelativistic effect governed by the underlying SPG symmetry.

In summary, we have developed a multipole classification scheme based on the IRREPs of the collinear SPG rather than the conventional MPG, which remains valid in the absence of SOC.
We derived the IRREPs of SPGs for collinear magnetic orders and identified the corresponding multipoles, enabling its unified classification. 
The resulting classification table provides symmetry-based criteria for determining which nonrelativistic responses are allowed or forbidden in a given collinear antiferromagnet.  
As a specific example, we demonstrated SOC-free spin current generation driven by elastic waves in the magnetic structure of CaMn$_2$Ge$_2$.
Our analysis thus offers a systematic, model-independent route to identifying nonrelativistic responses, and provides a foundation for exploring SOC-free functionalities in collinear antiferromagnets.

\begin{acknowledgments} 
This research was supported by JSPS KAKENHI Grants Numbers JP22H00101, JP22H01183, JP23H04869, JP23K03288, and by JST CREST (JPMJCR23O4) and JST FOREST (JPMJFR2366).
\end{acknowledgments}

\bibliography{18153.bib}
\bibliographystyle{jpsj}

\end{document}